\documentclass[useAMS,usenatbib]{mn2e}

\usepackage{graphicx}
\def\dM{\dot{M}}
\def\lcr{l_{\rm cr}}
\def\lmin{l_\star}
\def\fesc{f_{\rm esc}}
\def\Lesc{L_{\rm esc}}
\def\LZ{\bar{L}} 
\def\vZ{\bar{v}}
\def\tZ{\bar{t}}
\def\EZ{\bar{E}}
\def\OmZ{\bar{\Omega}}
\def\bOmZ{\bar{\bf \Omega}}
\def\Sesc{S_{\rm esc}}
\def\etaesc{\eta_{\rm esc}}
\def\rturn{r_{\rm turn}}
\def\aa{a_\star}
\title[Mini-discs around spinning black holes]
      {Mini-discs around spinning black holes}
\author[I. Zalamea and A. M. Beloborodov]
{I. Zalamea\thanks{E-mail: izalamea@phys.columbia.edu} and 
A. M. Beloborodov\thanks{Also at Astro-Space Center of Lebedev Physical 
Institute, Profsojuznaja 84/32, Moscow 117810, Russia} \\
Physics Department and Columbia Astrophysics Laboratory,
Columbia University, New York, NY 10027, USA\\
}
\begin{document}
\date{Accepted ---. Received ---; in original form ---}
\pagerange{\pageref{firstpage}--\pageref{lastpage}} \pubyear{---}
\maketitle
\label{firstpage}
\begin{abstract}
Accretion onto black holes in wind-fed binaries and in collapsars
forms small rotating discs with peculiar properties. Such ``mini-discs''
accrete on the free-fall time without help of viscosity and nevertheless
can have a high radiative efficiency. The inviscid mini-disc model was
previously constructed for a non-rotating black hole. We extend the model
to the case of a spinning black hole, calculate the structure and radiative
efficiency of the disc and find their dependence on the black hole spin.
If the angular momenta of the disc and the black hole are anti-aligned,
a hydrodynamic analog of Penrose process takes place.
\end{abstract}
%
%
\section{Introduction}
The mini-disc model was motivated by the estimate for angular momentum 
of accretion flows, $l$, in wind-fed X-ray binaries 
(Illarionov \& Sunyaev 1975; Shapiro \& Lightman 1976; 
Illarionov \& Beloborodov 2001). 
These systems happen to have $l\sim r_gc$, where
$r_g=2GM/c^2$ is the gravitational radius of the black hole. 
It is marginally sufficient to form a centrifugally supported disc.
Then a small disc can form, which is not supported centrifugally and 
instead accretes on the free-fall time (Beloborodov \& Illarionov 2001, 
hereafter BI01).
The mini-disc accretes so fast (super-sonically) that the effects of viscosity
can be neglected. 
A similar disc may form inside collapsing stars (Lee \& Ramirez-Ruiz 2006;
Beloborodov 2008).

The mini-disc can be thought of as a caustic in the equatorial plane of a
rotating accretion flow. It absorbs the feeding infall, and this 
interaction releases energy, making the accretion radiatively efficient.
With increasing angular momentum, the size of the disc grows up to $14r_gc$,
and at this point the centrifugal barrier stops accretion, so that it 
can proceed only on a viscous timescale.
Thus, the mini-disc model fills the gap between two classical regimes of 
accretion
  --- spherical ($l<r_gc$, Bondi 1952) and standard accretion disc
($l\gg r_gc$, Shakura \& Sunyaev 1973) --- and is qualitatively different 
from both.

The calculations of BI01 were limited to the case of a Schwarzschild black
hole. In the present paper we study the mini-disc around Kerr black holes.
The model is constructed under the following assumptions:

(i) The flow is axially symmetric. We assume that the rotational axes of 
the accretion flow and the black hole are aligned (or anti-aligned).

(ii) The flow is symmetric under reflection about the equatorial plane.
The symmetric streamlines collide in the equatorial plane and form a 
ring-like caustic around the black hole.

(iii) The flow falls freely (ballistically) from a large radius $r\gg r_g$
until it hits the caustic; its pressure is negligible everywhere except in 
the mini-disc.
This assumption is valid if the flow is cooled efficiently
(by radiation in X-ray binaries or by neutrinos in collapsars).
The heat released in 
the shocks that accompany the disc-infall interaction 
is assumed to be quickly radiated away, so that the shocks
stay near the disc plane, forming a ``sandwich''. 
The validity of this assumption is discussed in BI01
for the case of X-ray binaries and in Beloborodov (2008) for collapsars.

(iv) The flow is quasi-steady: its accretion rate and angular momentum
remain constant on the timescale of accretion through the mini-disc
(which is comparable to the free-fall time from $r\sim 10r_g$).

The paper is organized as follows. Section~2 describes the parabolic 
ballistic infall in Kerr metric and its ring caustic in the equatorial plane.
In section~3 we write down the equations that govern the gas motion in the 
caustic (the mini-disc) and solve the equations numerically. 
In section~4 we calculate the total luminosity of the disc observed at 
infinity, taking into account the light capture into the black hole.


\section{Supersonic infall with angular momentum}

The spacetime of a black hole of mass $M$ and angular momentum $J$ is 
described by the Kerr metric. In Boyer-Lindquist coordinates
$(ct,r,\theta,\phi)$ the metric is given by 
\begin{eqnarray}\label{eq1} \nonumber
 g_{ij}dx^{i}dx^{j}&=&-\left(1-\frac{r_{g}r}{\rho^{2}}\right)(cdt)^{2}-\frac{2r_{g}ar}{\rho^{2}
           }\sin^{2}\theta (cdt)d\phi\\ \nonumber
          & &+ \frac{\rho^{2}}{\Delta}dr^{2}+\rho^{2}d\theta^{2}+\frac{\sin^{2}\theta}{\rho^{2}}\\
                  & & + \left[(r^{2}+a^{2})^{2}-a^{2}\Delta\sin^{2}\theta\right]d\phi^{2}, 
\end{eqnarray}
\begin{eqnarray}\label{eq1}
 \rho^{2}&=&r^{2}+a^{2}\cos^{2}\theta,\\
   \Delta&=&r^{2}-r_{g}r+a^{2}.
\end{eqnarray}
The spin parameter of the black hole $a=J/Mc$ has dimension of cm and
must be in the interval $|a|\leq r_g/2=GM/c^2$. Throughout the paper
we shall also use the dimensionless parameter $\aa=ac^2/GM$, $|\aa|\leq 1$.

We assume that the gas infall forms at a large radius $r\gg r_g$, where it is 
efficiently cooled and, like dust, begins to fall freely
towards the black hole. 
A streamline of this 
ballistic infall is determined by three 
integrals of motion: specific angular momentum $l$, its 
projection $l_z$ on the spin axis of the black hole, and specific orbital 
energy $E\approx c^2$ (the infall is nearly parabolic). 
The four-velocity of a parabolic free-fall in Kerr metric
is given by (e.g. Misner et al. 1973),
\begin{equation}\label{eq21}
 \rho^{2}\frac{dr}{d\tau}=\pm c\sqrt{R},
\end{equation}
\begin{equation}\label{eq22}
\rho^{2}\frac{d\theta}{d\tau}=\pm\sqrt{\Theta},
\end{equation}
\begin{equation}\label{eq23}
\rho^{2}\frac{d\phi}{d\tau}=\frac{1}{\Delta}
       \left(c r_{g} a r -a^{2}l_z\right)+\frac{l_z}{\sin^{2}\theta},
\end{equation}
\begin{equation}\label{eq24}
\rho^{2}\frac{dt}{d\tau}=\frac{1}{\Delta}
      \left[\left(r^{2}+a^{2}\right)^{2} - r_{g} a r \frac{l_z}{c}\right]
     -a^{2}\sin^{2}\theta,
\end{equation}
where $l_z=l\sin\theta_\infty$, $\theta_\infty$ is the asymptotic 
polar angle of a streamline at large $r$, and
\begin{eqnarray}
\nonumber
  R(r) & = & r_{g} r^{3}-\frac{l^{2}}{c^{2}}r^{2}\left(1-\frac{r_{g}}{r}\right)
             - 2 r_{g}a \frac{l}{c} r \sin\theta_{\infty} \\
       &   & + a^{2}\left(r_{g}r-\frac{l^{2}}{c^{2}}\cos^{2}\theta_{\infty}\right), \\
  \Theta(\theta) & = & l_z^{2}(\cot^{2}\theta_{\infty}-\cot^{2}\theta).
\end{eqnarray}
The ballistic accretion flow is completely specified by the distribution of 
its density and angular momentum on a sphere of a large radius $r\gg r_g$. 
As the gas approaches the black hole and 
develops a significant rotational 
velocity, the infall is deflected from pure radial motion, 
and its streamlines intersect in the equatorial plane. 
The radius of this collision is determined by the angular momentum 
of the colliding symmetric streamlines.

\subsection{Collision Radius}

\begin{figure}
\begin{center}
\includegraphics[width=3.2in]{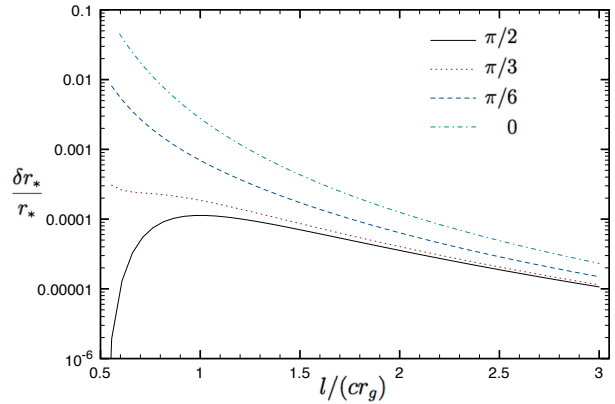}
\caption{
Fractional difference between the exact $r_\star$ and its approximation 
(eq.~\ref{rstar}) as a function of $l$. The curves start at $l_0=\lmin$ 
that corresponds to $r_\star$ at the horizon of the black hole.
All curves are plotted for maximally rotating black holes ($\aa=1$), 
when eq.~(\ref{rstar}) is least accurate.}
\label{fig_rsaccuracy}
\end{center}
\end{figure}

The streamline coming from an asymptotic direction 
$(\theta_{\infty},\phi_{\infty})$ with angular momentum 
$l(\theta_{\infty},\phi_{\infty})$ reaches the equatorial plane 
and collides with the symmetric streamline at the radius $r_\star$ 
defined by
\begin{equation}\label{eq31}
 -\int^{r_\star}_{\infty}\frac{dr}{\sqrt{R}}
    = \int^{\pi/2}_{\theta_{\infty}}\frac{c d\theta}{\sqrt{\Theta}}
    =\frac{\pi c}{2l}.
\end{equation}
We solved this equation for $r_\star$ numerically (it involves an elliptic 
integral on the left-hand side).
A good analytical approximation 
to $r_\star$ is given by
\begin{eqnarray}
\nonumber
  r_\star & \approx &\frac{l^{2}}{GM} -r_{g}\Big[ \frac{10-3\pi}{4}
             \left( \frac{l_{eff}}{l} \right)^{2} \\
      && -\frac{48-15\pi}{16} 
             \left( \frac{a c \cos\theta_{\infty}}{l} \right)^{2}\Big],
\label{rstar}
\end{eqnarray}
where $l_{eff}^{2} \equiv l^{2} + (a c)^{2}-2a c l\sin\theta_\infty$.
The accuracy of this approximation is shown in Figure~\ref{fig_rsaccuracy}.
It is better than 3 per cent when $r_\star$ is outside the horizon, 
for any set of parameters relevant for the mini-disc formation.
%

The collision radius is maximum for the streamlines near the equatorial plane
($\theta_\infty\rightarrow\pi/2$). This defines the outer radius of the disc,
\begin{equation}
  r_d=r_\star\left(\theta_\infty=\frac{\pi}{2}\right).
\end{equation}

\subsection{Non-intersection of streamlines above the equatorial plane}

The ballistic flow defines a mapping from the sphere 
$(\theta_\infty,\phi_\infty)$ to a sphere of radius $r$
(cf. BI01): $(\theta_{\infty},\phi_{\infty})\to(\theta(r),\phi(r))$.
The streamlines of the flow do not intersect before reaching the equatorial 
plane if the Jacobian of this mapping remains positive for all $r>r_\star$, 
i.e.
\begin{equation}
   {\cal J}=\textrm{Det} \left( \begin{array}{ccc}
  \displaystyle \frac{\partial\theta(r)}{\partial\theta_{\infty}} & 
  \displaystyle \frac{\partial\phi(r)}{\partial\theta_{\infty}} \\ &\\
  \displaystyle \frac{\partial\theta(r)}{\partial\phi_{\infty}} & 
  \displaystyle \frac{\partial\phi(r)}{\partial\phi_{\infty}}
                         \end{array} \right)>0.
\end{equation}
This condition is equivalent to 
$\partial\theta(r)/\partial\theta_{\infty}>0$. 
Using the relation between $\theta$ and $r$,
\begin{equation}
   \frac{\cos\theta(r)}{\cos\theta_{\infty}}
    =\cos\psi, \qquad \psi\equiv \frac{l}{c}\int^{\infty}_{r} \frac{dr}{\sqrt R}, 
\end{equation}
one finds that ${\cal J}>0$ if 
\begin{equation}\label{eq38}
    \frac{d\cos\theta(r)}{d\cos\theta_{\infty}} =\cos\psi
  -\cos\theta_{\infty}\sin\psi\frac{d\psi}{d\cos{\theta_{\infty}}}>0.
\end{equation}
This condition must be checked for a given distribution of angular momentum 
$l(\theta_\infty)$. In the numerical examples below we assume a 
distribution of the form
  \begin{equation}
  \label{eq:l}
    l(\theta_\infty)=l_{0}\sin\theta_{\infty},
  \end{equation}
(rigid-body rotation at infinity). We find that the streamlines do not
intersect before reaching the ring caustic in the equatorial plane.


\section{Disc Dynamics}\label{sec3}

We consider here only accretion flows that are asymptotically spherical
at $r\gg r_g$, i.e. we assume that the accretion rate at infinity 
is spherically symmetric, 
$d\dot{M}/d\Omega_\infty=const=\dot{M}_{\rm tot}/4\pi$. 
Streamlines that start at $\theta_\infty$ with angular momentum 
$l(\theta_\infty)$ reach the equatorial caustic at a radius
$r_\star(\theta_\infty)$.
The accretion rate through the disc at a radius $r$, $\dot{M}(r)$, 
equals the net accretion rate along the streamlines that 
enter the caustic outside $r$,
\begin{equation}
\label{eq:Mdot}
   \dot{M}(r)
             =\dot{M}_{\rm tot}\cos\theta_{\infty}(r),
\end{equation}
where $\theta_\infty(r)$ is the asymptotic polar angle of streamlines
that collide at radius $r$. We find it by inverting the function 
$r_\star(\theta_\infty)$.

Matter inside the disc moves horizontally with four-velocity 
$u^i=(u^t,u^r,0,u^\phi)$ and density $\rho$.
The disc is steady and axially symmetric, so $u^i$ and $\rho$ depend 
on $r$ only. Equations for $u^i(r)$ and $\rho(r)$ are derived
using the conservation laws for baryon number, energy and momentum.
These laws are expressed by the following general equations 
(e.g. Landau \& Lifshitz 1980),
\begin{equation}
 \label{eq:mass}
   \frac{1}{\sqrt{-g}}\partial_{i}(\sqrt{-g}\rho u^{i})=0,
\end{equation}
\begin{equation}
 \label{eq:stress}
   \frac{1}{\sqrt{-g}}\partial_{k}(\sqrt{-g}T^{k}_{i})
     =\frac{1}{2}\frac{\partial g_{kl}}{\partial x^{i}}T^{kl},
\end{equation}
where $T^{kl}=\rho c^{2}u^{k}u^{l}$ is the stress-energy tensor;
we assume that it is dust-like everywhere, i.e. neglect the internal
energy density, pressure, and magnetic fields compared with $\rho c^2$.
Above the disc, the stress-energy tensor is that of the ballistic infall.
The infall four-velocity and density just above the disc shall be 
denoted by $\hat{u}^i$ and $\hat{\rho}$, to distinguish them from the 
similar quantities inside the disc, $u^i$ and $\rho$.
Then the conservation laws (eqs.~\ref{eq:mass} and \ref{eq:stress}) give,
\begin{equation}
\label{eq:mass1}
  \frac{d\dot{M}}{dr}=-4\pi r^{2}\hat{\rho}\hat{u}^{\theta},
\end{equation}
\begin{equation}
\label{eq:stress1}
   \frac{d}{dr}\left(\frac{h}{r}T^{r}_{i}\sqrt{-g}\right)+2\sqrt{-g}\hat{T}^{\theta}_{i}
  =\frac{1}{2}\frac{h}{r}\sqrt{-g}\frac{\partial g_{kl}}{\partial x^{i}}T^{kl}.
\end{equation}
Here $\dot M=2\pi r h \rho u^{r}$, $h$ is the thickness of the disc, and
$\hat{T}^{kl}=\hat{\rho} c^{2}\hat{u}^{k}\hat{u}^{l}$ is the stress-energy 
tensor of the infall.
Using equation~(\ref{eq:mass1}) one finds from equation~(\ref{eq:stress1})
\begin{equation}
\label{eq:cons}
  \frac{d}{dr}(\dot{M}u_{i})-\hat{u}_{i}\frac{d\dot{M}}{dr}
  =\frac{1}{2}\frac{\partial g_{kl}}{\partial x^{i}}u^{k}u^{l}
   \frac{\dot{M}}{u^{r}}.
\end{equation}
For $i=\{t,\phi\}$ (conservation of energy and angular momentum)
the right-hand side of this equation vanishes. 
In particular, for $i=\phi$ this equation gives
\begin{eqnarray}
\label{eq:phi}
  \frac{d(\dot{M}u_{\phi})}{dr}=\hat{u}_{\phi}\frac{d\dot{M}}{dr},
\end{eqnarray}
\begin{eqnarray}   
  u_{\phi}(r)=-\frac{1}{\dot{M}(r)}\int_{r}^{r_{d}}\hat{u}_{\phi}
               \frac{d\dot{M}}{dr}dr.
\end{eqnarray}
For $l(\theta_\infty)$ given by equation~(\ref{eq:l})
we have 
$\hat{u}_{\phi}\equiv l_{z}=l^{2}/l_{0}$ and, using equation~(\ref{eq:Mdot}), 
we find 
\begin{equation}
\label{eq:uphi}
  u_\phi(r)=\frac{2}{3}l_{0}+\frac{l^{2}(r)}{3l_{0}}.
\end{equation}

\begin{figure}
\begin{center}
\includegraphics[width=3.2in]{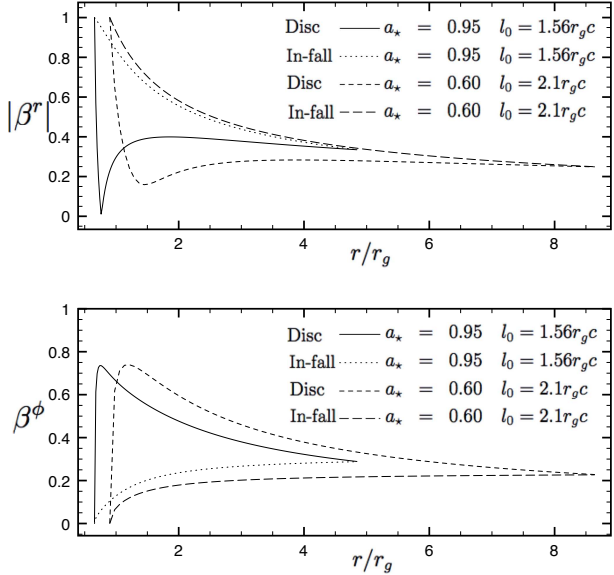}
\caption{{\it Upper panel:} 
solutions for the radial velocity measured by ZAMO (in units of $c$)
of the parabolic infall above the disc and the matter inside the disc. 
Two cases are shown: $\aa=0.6$, $l_0=2.1r_gc$ and $\aa=0.95$, $l_0=1.56r_gc$.
{\it Lower panel:} the corresponding azimuthal velocities of the infall
and the disc, measured by ZAMO.
}
\label{figVels}
\end{center}
\end{figure}

For $i=r$, equation~(\ref{eq:cons}) expresses conservation
of radial momentum. Its right-hand side does not vanish, and one needs 
to evaluate $\frac{\partial g_{kl}}{\partial r}u^{k}u^{l}$.
Substituting $u^{\phi}=(u_{\phi}-g_{t\phi}u^{t})/g_{\phi\phi}$, 
using $u_iu^i=-c^2$ 
and collecting terms one can write
\begin{equation}
\label{eq53}
   \frac{\partial g_{kl}}{\partial r}u^{k}u^{l}
   =A+B (u^{r})^{2}+Cu_{\phi}^{2}+Du_{\phi}u^{t},
\end{equation}
where $A,B,C$ and $D$ are functions of $r$ only,
\begin{eqnarray}\label{eq53b}
\nonumber
   A= -\frac{r_{g}c^{2} [r^{4}+a^{4}+2a^{2}r(r-r_{g})]}{r[a^{2}
      +r(r-r_{g})][r^{3}+a^{2}(r+r_{g})]},
\end{eqnarray}

\begin{eqnarray}
\nonumber
  B = 2\partial_{r}g_{rr}-\frac{a^{2}r(2r-3r_{g})}{[a^{2}
          +r(r-r_{g})][r^{3}+a^{2}(r+r_{g})]},
\end{eqnarray}

\begin{eqnarray}
\nonumber
  C= \frac{ r^{4}(2r-3r_{g})-2a^{4}r_{g}+a^{2}r(2r^{2}-3r_{g}r
          +3r_{g}^{2})}{[a^{2}+r(r-r_{g})][r^{3}+a^{2}(r+r_{g})]^{2}},
\end{eqnarray}
\begin{eqnarray}
  D= -\frac{2ar_{g}(3r^{2}+a^{2})}{[r^{3}+a^{2}(r+r_{g})]^{2}}.
\end{eqnarray}
Equation~(\ref{eq:cons}) for $i=r$ becomes
\begin{eqnarray}
\label{eq:rad1}\nonumber
  \frac{du^{r}}{dr}&=&\frac{\hat{u}^{r}-u^{r}}{\dot{M}}\frac{d\dot{M}}{dr}\\
&& +\frac{A+(B-2\partial_{r}g_{rr})(u^{r})^{2}
         +Cu_{\phi}^{2}+Du_{\phi}u^{t}}{2g_{rr}u^{r}}.
\end{eqnarray}
Finally, $u^t$ can be expressed in terms of $u^r$ and $u_\phi$ from
$u_iu^i=-c^2$,
\begin{eqnarray}
\label{eq:ut}\nonumber
  (u^{t})^{2}&=&1+\frac{r_{g}(r^{2}+a^{2})}{r[a^{2}+r(r-r_{g})]}
             +\frac{u_{\phi}^{2}/c^{2}}{a^{2}+r(r-r_{g})}\\
             &&+\frac{r^{4}+a^{2}r(r+r_{g})}{[a^{2}+r(r-r_{g})]^{2}}
                  \left(\frac{u^{r}}{c}\right)^{2}.
\end{eqnarray}
In equation~(\ref{eq:rad1}) all quantities are known functions of radius 
except $u^r$. We solve numerically this differential equation for $u^r(r)$.
Example solutions are shown in Fig.~\ref{figVels},
where we plot the radial and azimuthal velocities measured by ZAMO 
 (zero-angular-momentum observer at fixed r; Appendix~A gives the 
  transformation of 4-vectors to the ZAMO frame).

\begin{figure}
\begin{center}
\includegraphics[width=3.2in]{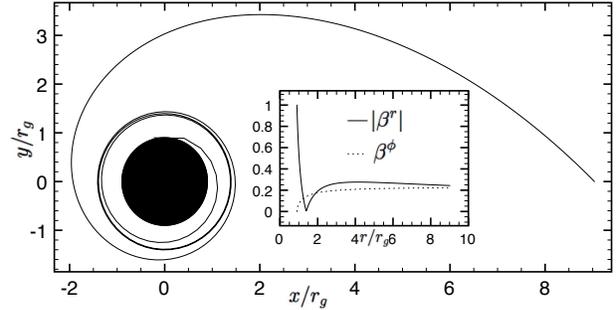}
\caption{Trajectory of disc matter in the critical case 
$l_0=2.14r_gc$ for a black hole with $\aa=0.6$. The insert 
shows the radial and azimuthal velocities of the disc measured by ZAMO.
$x$ and $y$ are the coordinates in the equatorial plane
defined by $x=r\cos\phi$, $y=r\sin\phi$.
}
\label{figTrajectoryCritical}
\end{center}
\end{figure}

\begin{figure}
\begin{center}
\includegraphics[width=3.2in]{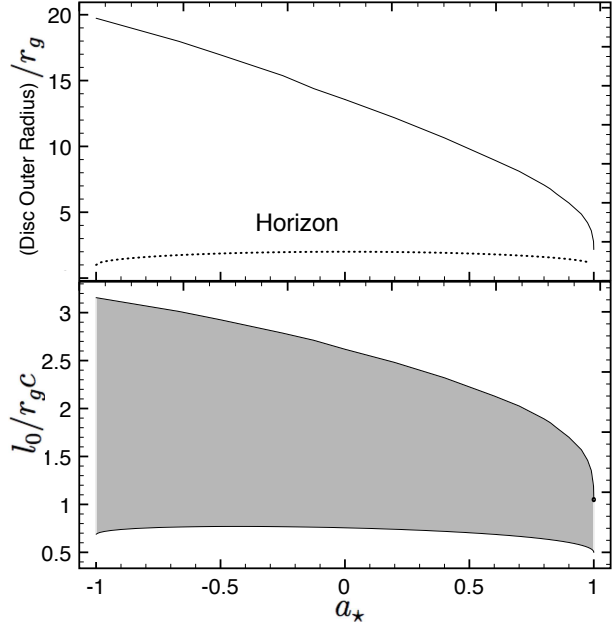}
\caption{
{\it Bottom panel:} the range of angular momenta $\lmin(\aa)<l_0<\lcr(\aa)$ 
that lead to the inviscid mini-disc regime (shaded region).
Accretion with $l_0<\lmin$ is quasi-spherical all the way into
the black hole; it does not form a caustic outside the horizon.
Accretion with $l_0>\lcr$ must proceed through a viscous, centrifugally 
supported disc.
  {\it Top panel:} maximum radius of the mini-disc ($l_0=\lcr$) as a function 
  of the black-hole spin parameter $\aa$ (solid curve). The radius of the 
  black hole $r_h(\aa)$ (eq.~\ref{eq:rh}) is shown by the dotted curve.
}
\label{figcriticallo}
\end{center}
\end{figure}

If the angular-momentum parameter of the flow, $l_0$, exceeds a critical 
value $\lcr$, accretion in the disc is stopped by the centrifugal barrier 
($u^r$ changes sign). For flows with $l_0<\lcr$ the radial velocity 
remains everywhere negative. As $l_0$ approaches $\lcr$ the trajectory of 
disc accretion makes more turns around the black hole 
(see Fig.~\ref{figTrajectoryCritical})
and at $l_0=\lcr$ it makes infinite number of turns. Similar behavior was 
found for mini-discs around Schwarzschild black holes; in this case 
$\lcr=2.62r_gc$ (BI01). For spinning black holes, $\lcr$ depends on $\aa$.
We have evaluated numerically $\lcr(\aa)$ and the corresponding maximum 
size of the mini-disc $r_d(\aa)$. The results are shown in 
Figure~\ref{figcriticallo}.

Figure~\ref{figcriticallo} also shows the minimum value of $l_0$, denoted 
by $\lmin$, that is required to form a disc outside the black-hole horizon,
\begin{equation} 
\label{eq:rh}
  r_h(\aa)=\frac{1+\sqrt{1-\aa^2}}{2}\,r_g.
\end{equation}
Accretion proceeds in the inviscid mini-disc regime when $\lmin<l_0<\lcr$.
This range shrinks with increasing $\aa>0$, and expands if $\aa<0$
(which means that the black hole and the accretion flow rotate in the 
opposite directions).

\begin{figure}
\begin{center}
\includegraphics[width=3.2in]{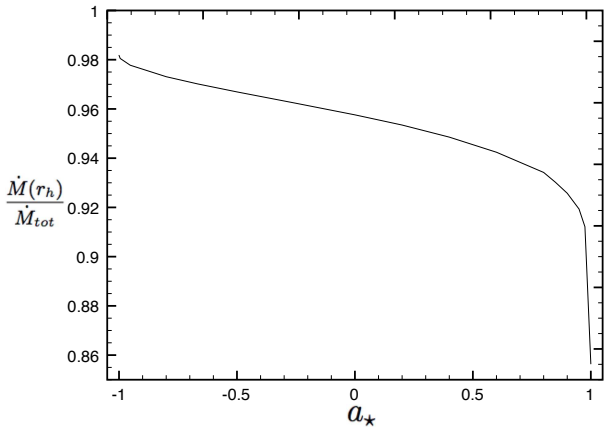}
\caption{Maximum mass fraction accreted through the inviscid mini-disc. 
This maximum corresponds to $l_0=\lcr$ and depends on the black-hole spin
$\aa$ (see the text).}
\label{figMatterRatio}
\end{center}
\end{figure}

Matter in the polar region of a quasi-spherical accretion flow falls 
directly into the black hole before crossing the equatorial plane. 
The mini-disc is formed in the other, equatorial part fo the flow
(cf. Fig.~1 in BI01). The boundary between these two accretion zones is 
determined by the condition $r_\star(l)=r_h$. Equation~(\ref{eq:Mdot}) 
gives the fraction of $\dM_{\rm tot}$ that accretes through the disc,
\begin{equation}
\label{eq:massfraction}
  \frac{\dot{M}(r_{h})}{\dot{M}_{\rm tot}}=\cos\theta_{\infty}(r_{h}),
\end{equation}
which depends on $l_{0}$ and $a$; it is maximum when $l_0=\lcr$.
The maximum fraction is shown in 
Figure~\ref{figMatterRatio}.

%


\section{Disc luminosity}

The mini-disc is a radiative caustic that converts kinetic energy of the 
acretion flow to radiation. Let $L(r)$ be the total luminosity 
produced by the disc outside radius $r$. The law of energy conservation 
gives an explicit expression for the luminosity,
\begin{eqnarray}\label{eq.luminosity}
   L(r)=\left[c^2+cu_t(r)\right]\dM(r),
\end{eqnarray}
where $c^2$ is the initial orbital energy of the 
accretion flow 
at infinity and $-cu_t(r)$ is the orbital energy of the disc material at 
radius $r$. 
It is found using the solution for $u^i(r)$ from section~3.
The radial distribution of luminosity is given by $dL/dr$,
and the total produced luminosity is $L=L(r_h)$.
This luminosity would be received at infinity if no radiation were
captured by the black hole. 

When the capture effect is taken into account,
the luminosity escaping to infinity may be written as 
\begin{equation}
  \Lesc=\int_{r_h}^{r_d}\frac{dL}{dr}\,\fesc(r)\,dr,
\end{equation}
where $\fesc(r)$ is the escaping fraction of radiation emitted at radius 
$r$; the fraction $1-\fesc$ is absorbed by the black hole. 
Assuming that the emission is approximately isotropic in the rest 
frame of the disc, we derive in Appendix~A
\begin{equation}
\label{eq:fesc}
   \fesc=\left(1-\displaystyle\frac{g_{t\phi}}
         {\sqrt{-\tilde g_{tt}g_{\phi\phi}}}
          \frac{\beta^{\phi}}{\gamma}\right)^{-1}
  \displaystyle \int_{S_{\rm esc}} 
   \frac{1-\displaystyle\frac{g_{t\phi}\OmZ^{\phi}}
                             {\sqrt{-\tilde g_{tt}g_{\phi\phi}}}}
        {\gamma^{4}(1-{\bf\beta}\cdot{\bOmZ})^{3}} \frac{d\OmZ}{4\pi},
\end{equation}
where ${\bf\beta}$ is the disc velocity (in units of $c$) measured by 
ZAMO (zero-angular-momentum observer at fixed $r$), 
$\gamma=(1-\beta^2)^{-1/2}$ and  
$\tilde{g}_{tt}= g_{tt}-g_{t\phi}^{2}/g_{\phi\phi}$. The integral is taken 
over the escape cone $\Sesc(r)$ --- all photon directions $\bOmZ$ 
(in the ZAMO frame) that lead to escape. The calculation of these cones is 
described in Appendix~A. 

\begin{figure}
\begin{center}
\includegraphics[width=3.2in]{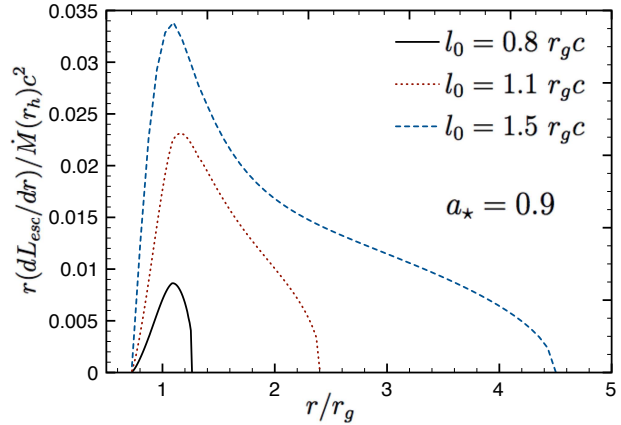}
\caption{Radial distribution of escaping luminosity for discs 
with $l_0/r_gc=0.8$, 1.1, and 1.5 around a black hole with spin parameter
$\aa=0.9$. The luminosity is normalized by the rate of rest-mass 
accretion through the disc, $\dot{M}(r_h)c^2$. }
\label{figLumDis09}
\end{center}
\end{figure}

The resulting $\Lesc(r)$ is found numerically.
Figure~\ref{figLumDis09} shows the radial distribution $d\Lesc/dr$ for 
mini-discs with three different $l_0$ around a black hole with spin 
parameter $\aa=0.9$. 
Most of the luminosity is produced in the region $r_g<r<1.4r_g$ where the 
infall velocity relative to the disc is large and hence a large energy is
released in the disc-infall interaction.

\begin{figure}
\begin{center}
\includegraphics[width=3.2in]{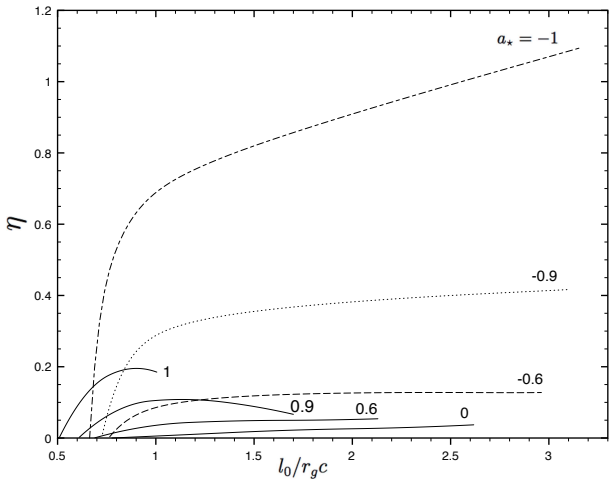}
\caption{Radiative efficiency as a function of $l_0$, ignoring the light 
capture into the black hole. Seven curves are plotted for black holes with 
different spin parameters $\aa$. Solid curves are used for $\aa\geqslant0$ 
and broken curves for $\aa<0$.
}
\label{figUnCefficiency}
\end{center}
\end{figure}

We define the radiative efficiency of the disc as the ratio of the total 
luminosity radiated to infinity, $\Lesc=\Lesc(r_h)$, to the rest-mass flux 
through the disc,
\begin{equation}\label{eq104}
  \etaesc=\frac{\Lesc}{\dot M(r_h)c^2}.
\end{equation}
If the light capture into the black hole is ignored, i.e.
$\Lesc$ is replaced by $L$, the efficiency is given by 
\begin{equation}
\label{eq:eta}
    \eta=\frac{L(r_{h})}{\dot M(r_h)c^2}=1+\frac{u_t(r_h)}{c}.\\ 
\end{equation}
We evaluated numerically the dependence of $\eta$ on the two parameters of 
the mini-disc $l_0$ and $\aa$ (Fig.~\ref{figUnCefficiency}). This dependence 
may be better understood if we express $u_t(r_h)$ in terms of $u^r(r_h)$ and 
$u_\phi(r_h)$ from $u_iu^i=-c^2$; then equation~(\ref{eq:eta}) yields
\begin{eqnarray}
\nonumber
   \eta &=&1-\left(1+\sqrt{1-\aa^{2}}\right)^{1/2}\frac{|u^{r}(r_{h})|}
                                                  {c\sqrt 2}\\ 
        & &-\aa\left(1+\sqrt{1-\aa^{2}}\right)^{-1/2}
           \displaystyle\frac{u_{\phi}(r_{h})}{r_{g}c}.
\label{eq:eta1}
\end{eqnarray}
If $\aa=0$ this equation simplifies to $\eta=1-|u^r(r_h)|/c$ and gives 
a monotonic dependence of $\eta$ on $l_0$: flows with larger angular momenta 
have smaller $|u^r(r_h)|$ (their radial motion is centrifugally decelerated) 
and higher $\eta$. 
In the case of a rotating black hole, equation~(\ref{eq:eta1}) has additional
terms which lead to a complicated dependence of $\eta$ on $l_0$ and $\aa$.
For instance, when $\aa=0.9$ the dependence of $\eta$ on $l_0$ is not 
monotonic (Fig.~\ref{figUnCefficiency}).

The efficiency $\eta$ is generally increasing with increasing spin of the
black hole. It is especially high for retrograde discs, which are described 
by the solutions with $\aa<0$. In this case, the second term on the right-hand 
side of equation~(\ref{eq:eta1}) is positive and can substantially increase
$\eta$. A remarkable feature of retrograde discs is that they can extract 
energy from the black hole via a hydrodynamic analog of Penrose process. 
This occurs if $\aa$ is close to $-1$; in such discs $u_t(r_h)<0$ and 
$\eta>1$. The maximum $\eta=1.1$ is reached when $\aa=-1$ and 
$l_0=\lcr(-1)=3.15r_gc$. 

\begin{figure}
\begin{center}
\includegraphics[width=3.2in]{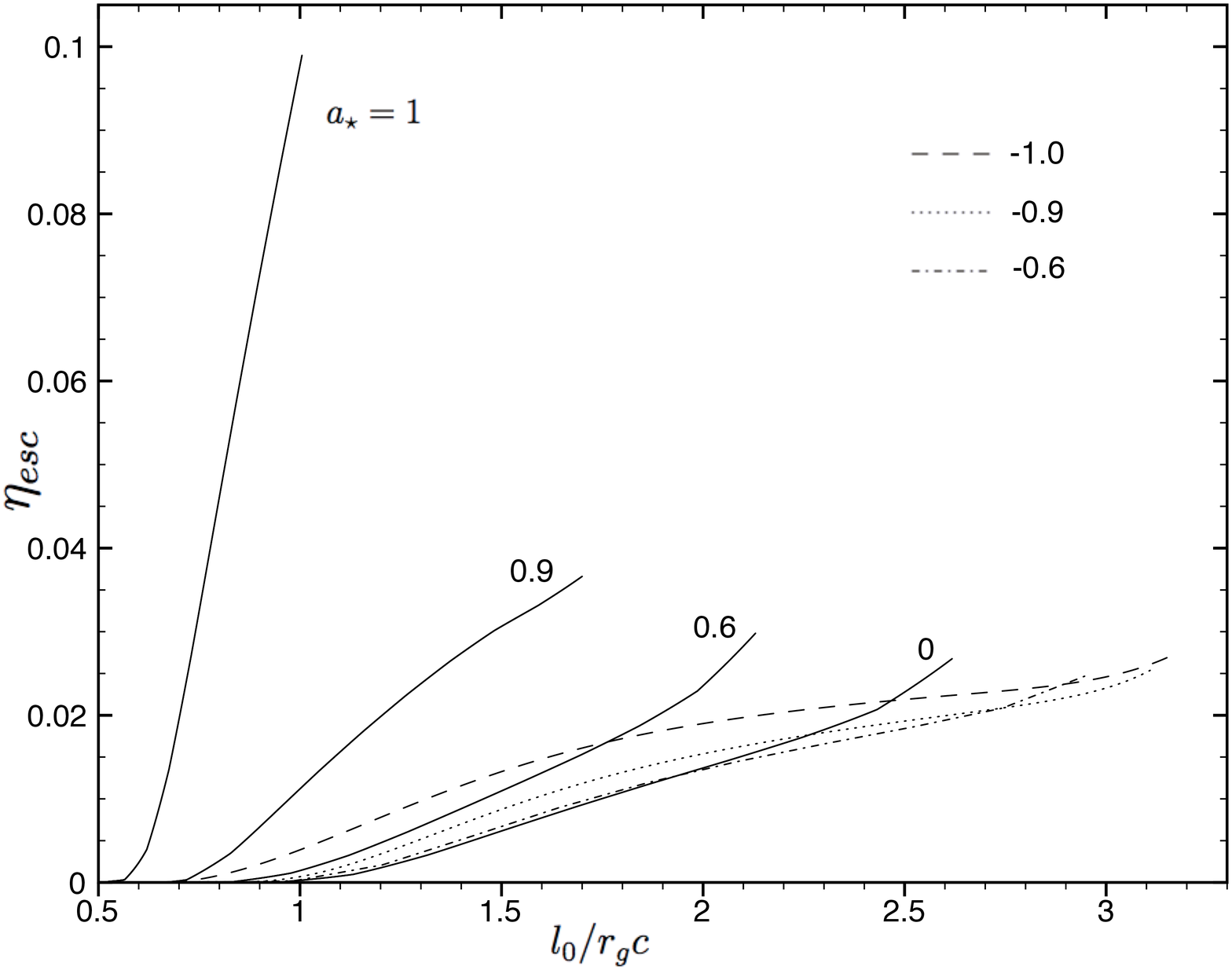}
\caption{Same as Fig.~\ref{figUnCefficiency} but taking into account the 
suppression of radiative efficiency due to light capture into the black hole.
}
\label{figCefficiency}
\end{center}
\end{figure}

Next, we evaluated numerically the efficiency $\etaesc$ that takes into 
account the capture of the produced radiation into the black hole 
(Fig.~\ref{figCefficiency}). This effect greatly reduces the observed
luminosity.
The reduction is especially significant for retrograde discs because their
radiation is Doppler-beamed in the direction opposite to the black hole 
rotation, so most of their radiation misses the escape cones shown in 
Figure~\ref{figEscapeBounds}. The resulting efficiency $\etaesc$ is highest
for prograde discs around maximally rotating black holes ($\aa=1$). The 
maximum $\etaesc$ is close to 10 per cent.

\section{Conclusions}
                                                                                
In our model, the mini-disc is completely described by two parameters:
the maximum angular momentum of the accreting gas $l_0$ and the spin 
parameter of the black hole $\aa$. $l_0=0$ corresponds to spherical 
accretion. The inviscid mini-disc forms when $l_0$ is in the range 
$\lmin(\aa)<l_0<\lcr(\aa)$ shown in Fig.~\ref{figcriticallo}.

When the black hole rotates in the same direction as the accretion flow 
($\aa>0$), this range becomes smaller compared to the Schwarzschild case 
and the maximum possible size of the mini-disc is reduced. 
For example, for the typical $\aa\sim 0.9$ expected in collapsars we find 
$\lcr\approx 1.5 r_gc$, which corresponds to a disc of radius 
$r_d\approx 6r_g=12GM/c^2$. The mini-disc model may 
describe collapsars at the early stage when the disc grows from the black 
hole horizon $r_h$ to $\sim 12GM/c^2$; then it must switch to the standard 
viscous regime.

We have calculated the radiative efficiency of mini-discs taking into 
account the effect of photon (or neutrino) capture into the black hole 
(Fig.~\ref{figCefficiency}). The efficiency is maximum when the mini-disc 
has its maximum size, near the transition to the viscous-disc regime. 
For prograde discs around rapidly rotating black holes, the efficiency 
approaches 0.1, which is $\sim 3$ times higher than for non-rotating 
black holes. 

We have also studied the case of retrograde discs, where the disc and 
the black hole are counter-rotating ($\aa<0$). Such discs can form
in wind-fed X-ray binaries where gas accretes with alternating angular
momentum due to fluctuations and `flip-flop' instability (e.g. Shapiro 
\& Lightman 1976; Blondin \& Pope 2009 and refs. therein). We find that
a hydrodynamic analog of Penrose process works in such discs if $\aa$
is close to $-1$, i.e. the black hole is close to maximum rotation.
Then the black hole accretes matter with negative orbital energy,
which means that energy is extracted from the black hole. 
However, if the extracted energy is radiated quasi-isotropically in the 
disc rest-frame, most of the produced radiation ends up inside the black 
hole, and the escaping luminosity is suppressed.


                                                                                
\appendix
                                                                                
\section[]{Calculation of the escaping luminosity}
\label{app:Sr}

\subsection{Angular distribution of emission in the local ZAMO frame} 

Consider an infinitesimal element of the disc, a ring of radius $r$
and thickness $\delta r$. It produces luminosity 
\begin{equation}
\label{eq:dL}
  \delta L=\frac{dL}{dr}\,\delta r
                 =\frac{d}{dr}\left[(c^2+cu_t)\dot{M}\right]\,\delta r.
\end{equation}
The ring emits photons in all directions, which will be parameterized
by unit 3D vector $\bOmZ$ in the local frame of ZAMO 
  (zero-angular-momentum observer at fixed $r$, 
see e.g. Misner et al. 1973). 
Then $\delta L$ may be written as
\begin{equation}
\label{eq:dL1}
   \delta L=\int_{4\pi} \frac{d(\delta L)}{d\OmZ}\,d\OmZ
                  =\int \frac{d\tZ}{dt} \frac{E}{\EZ}
                        \frac{d(\delta\LZ)}{d\OmZ}\,d\OmZ.
\end{equation}
Here $\delta \LZ$ is the luminosity of the ring measured by ZAMO and $\tZ$ 
is the proper time of ZAMO. $E/\EZ$ is the ratio of photon energy measured 
at infinity, $E=-cv_t$, and measured by ZAMO, $\EZ=-c\vZ_t=c\vZ^t$; 
this ratio depends on the emission direction $\bOmZ$. It is found 
from the Lorentz transformation of the photon 4-velocity from the coordinate
basis to the (orthonormal) ZAMO basis, 
\begin{equation}\label{eq:transf}
  \left(\begin{array}{c}
                 \vZ^t    \\
                 \vZ^\phi
        \end{array}
  \right) = \left( \begin{array}{cc}
     \sqrt{-\tilde{g}_{tt}}        & 0 \\
     g_{t\phi}/\sqrt{g_{\phi\phi}} & \sqrt{g_{\phi\phi}}
                   \end{array}
            \right)
            \left(\begin{array}{c}
                     v^t\\
                     v^\phi
                  \end{array}
            \right),
\end{equation}
\begin{eqnarray}
\label{eq:transf1}
   \vZ^r=\sqrt{g_{rr}}v^{r},   \qquad 
   \vZ^\theta=\sqrt{g_{\theta\theta}}v^{\theta},
\end{eqnarray}
where $\tilde{g}_{tt}= g_{tt}-g_{t\phi}^{2}/g_{\phi\phi}$. 
Then one finds,
\begin{equation}
\label{eq:EE}
   \frac{E}{\EZ}=\frac{v_t}{\vZ_t}
   =\sqrt{-\tilde g_{tt}}-\frac{g_{t\phi}}{\sqrt{g_{\phi\phi}}}\OmZ^{\phi}.
\end{equation}
The ratio $d\tZ/dt$ appearing in equation~(\ref{eq:dL1})
equals $(-\tilde{g}_{tt})^{1/2}$.

The angular distribution of luminosity measured by ZAMO, 
$d(\delta\LZ)/d\OmZ$, is related to the angular distribution
of luminosity in the rest frame of the accreting gas of the disc,
$d(\delta L_c)/d\Omega_c$, by the Doppler transformation 
(see e.g. Rybicki \& Lightman 1979),
\begin{equation}
\label{eq:Dop}
  \frac{d(\delta\LZ)}{d\OmZ}
 =\frac{1}{\gamma^4(1-{\bf\beta}\cdot\bOmZ)^3}
  \frac{d(\delta L_c)}{d\Omega_c},
\end{equation}
where ${\bf\beta}$ is the disc velocity (in units of $c$) measured by ZAMO.
We assume that emission is approximately isotropic in the gas frame,
i.e. $d(\delta L_c)/d\Omega_c=\delta L_c/4\pi$.
Substitution of equations~(\ref{eq:EE}) and (\ref{eq:Dop}) to 
equation~(\ref{eq:dL1}) gives,
\begin{equation}
\label{eq:dL2}
  \delta L=\frac{\delta L_c}{4\pi} \left(-\tilde{g}_{tt}\right)^{1/2}
    \int_{4\pi}\frac{[\sqrt{-\tilde g_{tt}}
                 -(g_{t\phi}/\sqrt{g_{\phi\phi}})\OmZ^{\phi}]}
        {\gamma^4(1-{\bf\beta}\cdot\bOmZ)^3}\,d\OmZ.
\end{equation}
The expression for escaping luminosity $\delta\Lesc$ is similar except that 
the integral is taken over the escape cone $\Sesc$ rather than $4\pi$.
Therefore, the escaping fraction is given by 
\begin{eqnarray}
\nonumber
  \fesc\equiv\frac{\delta\Lesc}{\delta L}
     =\left(\int_{4\pi}\frac{[\sqrt{-\tilde g_{tt}}
              -(g_{t\phi}/\sqrt{g_{\phi\phi}})\OmZ^\phi]}
             {\gamma^4(1-{\bf\beta}\cdot\bOmZ)^3}\,d\OmZ\right)^{-1} \\
       \times \int_{\Sesc}\frac{[\sqrt{-\tilde g_{tt}}
              -(g_{t\phi}/\sqrt{g_{\phi\phi}})\OmZ^\phi]}
             {\gamma^4(1-{\bf\beta}\cdot\bOmZ)^3}\,d\OmZ.
\nonumber
\end{eqnarray}
Evaluating the integral over $4\pi$ we obtain equation~(\ref{eq:fesc}).

\subsection{Escape cones}

The photon direction in the ZAMO frame, $\bOmZ$, can be described by 
two angles $\alpha$ and $\varphi$,
\begin{eqnarray}
\label{eq:Omr}
 \OmZ^{r}=\frac{\vZ^r}{\vZ^t}               &=& \cos{\alpha}, \\ 
\label{eq:Omth}
\bar \Omega^{\theta}=\frac{\vZ^\theta}{\vZ^{t}} &=& \sin{\alpha}\sin{\varphi}, \\
\label{eq:Omphi}
\bar \Omega^{\phi}=\frac{\vZ^\phi}{\vZ^t}       &=& \sin{\alpha}\cos{\varphi}.
\end{eqnarray}
Using the relations between $\vZ^i$ and $v^i$ (eqs.~\ref{eq:transf} and 
\ref{eq:transf1}), we express $\bOmZ$ in terms of $v^i$,
\begin{eqnarray}
  \cos{\alpha}=\sqrt{-\frac{g_{rr}}{\tilde{g}_{tt}}}\frac{v^r}{v^t}
              =\frac{\sqrt{r^4+a^2r(r+r_g)}}{a^2+r(r-r_g)}
               \frac{v^r}{v^t},\\
  \sin\alpha\sin\varphi=\sqrt{-\frac{g_{\theta\theta}}{\tilde{g}_{tt}}}
                         \frac{v^\theta}{v^t}
                       =\sqrt{\frac{r^4+a^2r(r+r_g)}{a^2
                        +r(r-r_g)}}\frac{v^\theta}{v^t},
\end{eqnarray}
where all metric coefficients have been evaluated at the equatorial plane
at the emission radius $r$.

The photon 4-velocity $v^i$ is expressible in terms of four integrals of 
motion in Kerr metric: energy $E$, total angular momentum $L$, its 
projection $L_z$, and Carter integral $\Gamma$. Using these expressions 
(see e.g. Chandrasekar 1982) and choosing the affine parameter along the 
photon worldline so that $E=c^2$, we find $v^r/v^t$ and $v^\theta/v^t$ as 
functions of $r$, $\Gamma$ and $L_z$, and obtain
\begin{eqnarray}
\nonumber
  \cos^2\alpha &=& \frac{r^4+a^2r(r+r_g)}
     {\left[ r^4+a^2r(r+r_g) - r_g a L_z r \right]^2} \\
\nonumber
  &\times& \left[ r^4+r^2\left(\Gamma+a^2\right)
       +r r_g\left(a^2-\Gamma-2aL_z\right)\right. \\ 
  & &  \left.+a^2\left(\Gamma+L_z^2\right)\right],
\label{eq:1}
\end{eqnarray}
                                                                                
\begin{eqnarray}\nonumber
  \sin^2\alpha\,\sin^2\varphi &=& -\left(\Gamma+L_z^2\right)
                                 \left( r^2+a^2 - r_g r\right) \\ 
\label{eq:2}
   & & \times \frac{\left[r^4+a^2r(r+r_g)\right]}
                   {\left[ r^4+a^2r(r+r_g)-r_gaL_zr \right]^2}.
\end{eqnarray}
Equations (\ref{eq:1}) and (\ref{eq:2}) can be solved for
$L_z$ and  $\Gamma$ for given $r$, $\alpha$ and $\varphi$.

The fate of a photon is determined by the equation of radial motion
for null geodesics (e.g. Chandrasekar 1982),
\begin{eqnarray}
\nonumber
  \left(r^2+a^2\cos^2\theta\right)^2\left(v^r\right)^{2}
  =\left(r^2+a^2-r_gr\right)\left(\Gamma-a^2E^2\right) \\
   +\left(r^2+a^2\right)^2E^2-2r_gaEL_zr+a^2L_z^2,
\end{eqnarray}
where $E=c^2$ with our choice of affine parameter along photon worldline.
The possible turning points $\rturn$ are found from the condition $v^r=0$.
Let $r_0$ be the initial radial position of the photon. 
Two cases are possible:
(i) $v(r_0)>0$, the photon escapes if there are no $\rturn>r_0$. 
(ii) $v(r_0)<0$, the photon escapes if there is at least one turning 
point such that $r_h<\rturn<r_0$ and no $\rturn>r_0$.

\begin{figure}
\begin{center}
\includegraphics[width=3.2in]{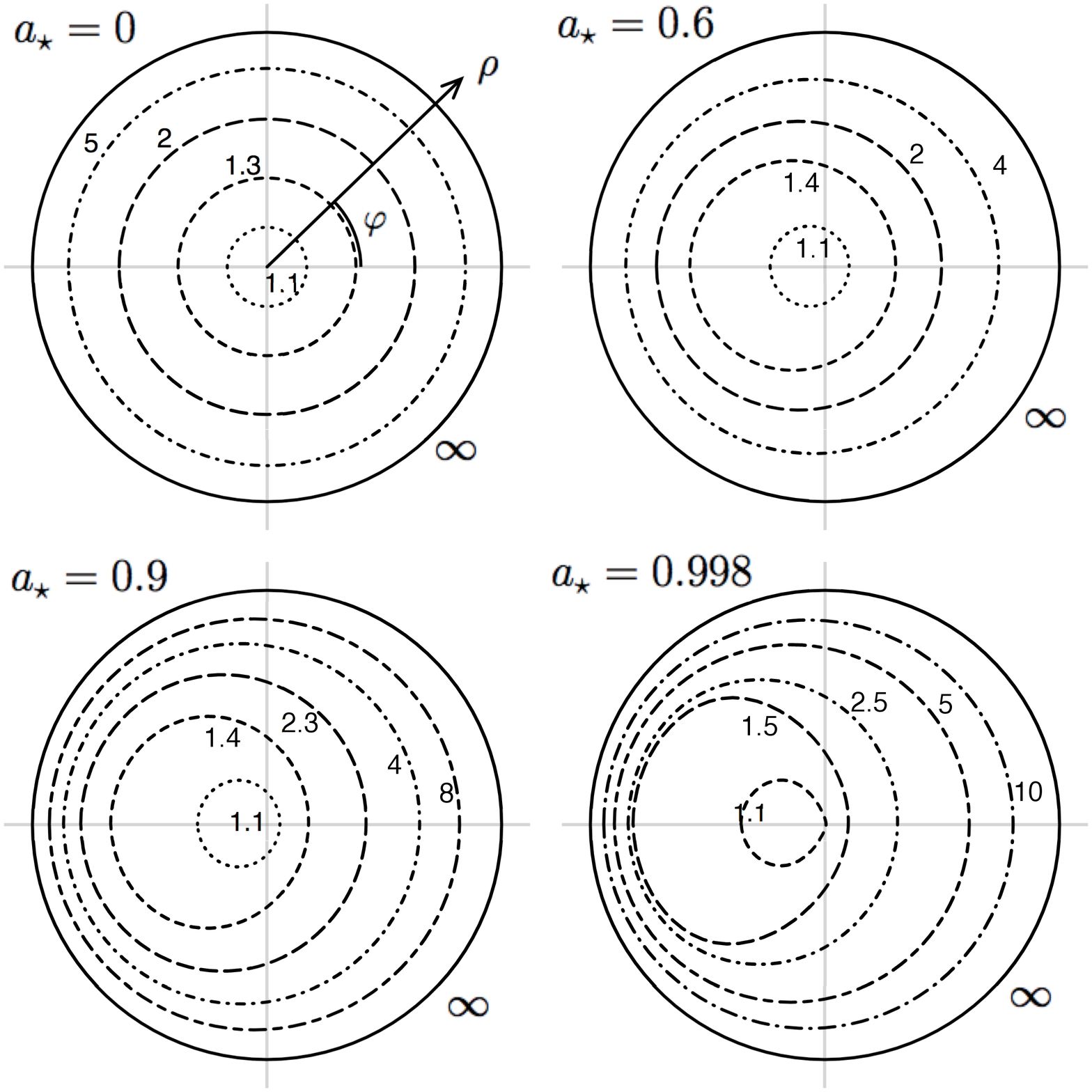}
\caption{Escape cones $\Sesc$ on the ZAMO sky. 
The photon direction $\bOmZ$ is specified by two angles $\alpha$ 
and $\varphi$ (eqs.~\ref{eq:Omr}-\ref{eq:Omphi}).
The figure uses polar coordinates $(\rho,\phi)$ with $\rho=\alpha/\pi$
to represent all possible photon directions. The origin of the diagram
$\rho=0$ corresponds to the radial direction away from the black hole 
(such photons always escape) and the unit circle $\rho=1$ (thick black 
curve) corresponds to the radial direction into the black hole 
(such photons are captured). The colour curves show the boundary of the 
escape cone for five emission radii; the emission radius is indicated 
next to the curves, in units of the horizon radius $r_h$.
The figure presents four such diagrams calculated for black holes
with spin parameter $\aa=0$, 0.6, 0.9, and 0.998.
}
\label{figEscapeBounds}
\end{center}
\end{figure}

For any given direction $\bOmZ$, we find $L_z$ and $\Gamma$ from 
equations~(\ref{eq:1}) and (\ref{eq:2}), then determine the roots 
$\rturn$ of equation $v^r(r)=0$ and check the escape conditions.
All escaping directions form the ``cone'' $\Sesc$ on the sky of ZAMO,
which is found numerically. 
  Figure~\ref{figEscapeBounds} shows the escape cones for five 
  emission radii $r_0$ and four values of $\aa$.

\end{document}